\documentstyle[preprint,aps]{revtex}
\begin{document}

\title{Spectroscopy of Local Density of States Fluctuations in a Disordered
Conductor}

\author{T. Schmidt, R. J. Haug\cite{h}, Vladimir I. Fal'ko\cite{f},
and K. v. Klitzing}
\address{Max-Planck-Institut f\"ur Festk\"orperforschung, Heisenbergstr. 1,
70569 Stuttgart, Germany}

\author{A. F\"orster and H. L\"uth}
\address{Institut f\"ur Schicht- und Ionentechnik,
Forschungszentrum J\"ulich GmbH, Postfach 1913,\\
52428 J\"ulich, Germany}

\date{September 8, 1995}
\maketitle

\begin{abstract}
The local density of states of a degenerate semiconductor is
investigated at low magnetic fields by resonant tunneling through a
discrete localized electron level.
Fine structure in the tunneling current provides a
temperature insensitive image of mesoscopic fluctuations
of the local density of states below the Fermi level.
The fluctuations are demonstrated to originate from quantum interference of
diffusive electron waves in the three-dimensional disordered system.
\end{abstract}

Quantum mechanics is well known to manifest itself in the low-temperature
transport properties of disordered conductors.
Weak localization corrections to the conductivity \cite{ber84} and
universal conductance fluctuations \cite{was86} are the phenomena
most explored experimentally in disordered metals.
Their theoretical understanding is based on the interference of
diffusive electron waves \cite{ber84,alt91} which
on one hand, suppresses quantum diffusion and, on the other hand,
makes the transport properties sample-specific since the interference
pattern depends on the configuration of disorder.
Universal conductance fluctuations are entirely determined by the properties
of electron states at the Fermi level.
According to existing theory, however, fluctuations in the spectrum of
levels as well as the structure of wavefunctions are reflected in full detail
only in the density of states
\cite{alt86} and, especially, the local density of states
(LDOS) \cite{ler88} of disordered systems.
Recently, the first LDOS observation was realized
by employing scanning-tunneling microscopy to image the influence of
an external modulation on the two-dimensional surface bands
of noble metals \cite{cro93}.
In contrast, the experimental study of LDOS fluctuations in
disordered three-dimensional conductors was impossible
until now due to the lack of an appropriate spectrometer.

In this letter, we analyze the LDOS of a three-dimensional (3D)
degenerate semiconductor by means of resonant tunneling through a
discrete localized electron state.
Resonant tunneling is known as a spectroscopic tool for probing
the properties of electronic systems~\cite{siv94}.
Recently, it has been proposed to employ resonant tunneling
through impurity states for studying the LDOS in the contacts
of metal-insulator-metal junctions~\cite{ler92}.
The idea of our experiment is slightly different as illustrated
by the sketch in Fig. \ref{sketch}.
Electrons tunnel from the heavily-doped disordered
emitter of a double-barrier heterostructure through
the energetically-lowest level $S$ in the quantum well which
serves as {\em spectrometer} because it is narrow in energy and
localized in space. Such discrete localized states were
recently shown to have a defect-related origin \cite{del92}.
The tunneling current $I$ is determined by the transparency of the
thick emitter barrier as well as by the {\em local tunneling
density of states} $\nu$ in the emitter at the position of the
spectrometer, $I\propto \nu$. As a bias voltage $V$ is applied,
the spectrometer scans the LDOS below the Fermi level
as a function of energy.
The measured tunneling current exhibits an irregular modulation
which reflects exclusively LDOS fluctuations, $\delta I\propto\delta\nu$,
since the emitter-barrier transparency is a smooth function of energy.
The observed fluctuation pattern is analyzed in detail
as a function of bias voltage and magnetic field.

The experiment was performed using a strongly asymmetric double-barrier
heterostructure grown by molecular-beam epitaxy on
$n^+$-type GaAs substrate.
The structure consists of a 10 nm wide GaAs quantum
well sandwiched between two Al$_{0.3}$Ga$_{0.7}$As barriers of
5 and 8 nm width (top and bottom barrier). The undoped active
region is separated from 300 nm thick
GaAs contact layers doped with Si to
$4\times 10^{17}$ cm$^{-3}$ by 7 nm wide GaAs spacer layers.
{}From this material we fabricated a pillar with a diameter of 500 nm
supplied with ohmic top and bottom contacts.
The current and the differential conductance of the device were
independently measured in a dilution refrigerator. A lock-in amplification
technique was used with an ac excitation voltage of 15 $\mu$V amplitude and
13 Hz frequency superimposed on the dc bias voltage applied to the top
contact.

Figure \ref{sketch} (a) shows the low-bias part of the
current-voltage characteristic $I(V)$ measured at
the base temperature of $T\approx 30$ mK of the dilution refrigerator.
Two steps due to resonant tunneling through discrete levels
are clearly observed. The first step in the current occurs at the
voltage at which the energetically-lowest discrete state $S$
in the quantum well crosses the electrochemical potential
$\mu_E$ in the emitter contact, see sketch on top of Fig. \ref{sketch}.
The voltage scale is related to the energy scale by $E=\alpha eV+const$
with $\alpha\approx 0.5$ denoting the voltage-to-energy
conversion coefficient. The second step arises when the next,
energetically higher discrete state becomes available for transport.

The most striking feature of the data is the reproducible oscillatory
fine structure which is superimposed on the current plateau
between the two current steps. Indications of similar fluctuations
were recently observed by several groups \cite{su92,gei94,des94,mcd95}.
The current plateau is formed by electrons which tunnel from
below the emitter Fermi level through the lowest discrete state
in the quantum well.
The tunneling current depends on the transparencies of the barriers and
the density of states in both contacts.
In order to eliminate the influence of the collector-barrier
tunneling process, we designed the double-barrier heterostructure
strongly asymmetric. The transparency of the thick emitter barrier
is by orders of magnitude lower than that of
the collector barrier. Hence, it is the emitter-tunneling
process which determines the magnitude of the
current. The observed fine structure represents thus a direct image of
fluctuations of the density of states in the emitter contact,
$\delta I\propto \delta\nu$. This image
is recorded by using the energetically-lowest discrete state in the
quantum well as spectrometer. The finite slope of the current plateau is
due to the smooth bias-voltage dependence of the transparency of the
emitter barrier \cite{sch95a}.

The fine structure is even more pronounced in the simultaneously measured
differential conductance $G=dI/dV$ plotted in
Fig. \ref{sketch} (b) for $T\approx 30$ mK and $T=1.0$ K.
Considering differential conductance instead of current eliminates the
finite slope of the current plateaus.
Large resonances in the differential conductance originate from
the current steps in the $I(V)$ characteristic. The amplitude of these
main resonances is sensitive to thermal broadening of the Fermi
distribution since they are due to electrons tunneling from the
Fermi level. In contrast, the fine structure is practically
temperature independent. This independence complies with the assumption of
resonant tunneling from below the Fermi level where at
$T=1.0$ K all emitter states are still occupied.

In order to investigate the quantum origin of the density of states
fluctuations, we employed a magnetic field to vary the interference
conditions for diffusive electron waves in the emitter contact.
Figure \ref{cult} shows the differential conductance
$G$ as a function of both the bias voltage $V$ and a magnetic field $B$
parallel to the current flow.
The position of the main resonances in bias is up to $B=2$ T
independent of the magnetic field, which demonstrates that the
corresponding discrete electron states in the quantum well
are strongly localized in space. Measurements up to higher fields
yield a radius of $r\sim 10$ nm for the energetically-lowest state
(in analogy to Ref. \cite{sch95a}) which suggests
that the spectrometer is a single impurity within the
double-barrier region. The spectral resolution of our experiment is
deduced from the width of the first conductance resonance as
$\gamma\approx 0.2$ meV. Since the {\em spectrometer} is strongly
localized, it is the {\em local density of states} at the lateral
position of this discrete level which determines the oscillatory fine
structure between the two main resonances.
It is however important to realize that the differential
conductance displays in contrast to the current not directly
density of states fluctuations but their derivative with respect
to energy, $\delta G\propto \delta(d\nu/dE)$. The fine structure
shows a strong irregular dependence on magnetic field which
is the subject of the following statistical analysis.

The analysis of the fluctuation pattern is based on the evaluation of
correlation functions
in the intervals $V=25-30$ mV and $B=0-2$ T.
Fig. \ref{stat1} (a) represents the autocorrelation
function $\left<K_{\Delta B}\right>_V=\left<\left<\delta G(B)\delta
G(B+\Delta B)\right>_B/{\rm v}{\rm a}{\rm r}_BG\right>_V$ as a function of
magnetic field with $\delta G(B)=G(B)-\left<G(B)\right>_B$ and
var$_BG=\left<\delta G^2(B)\right>_B$. The symbols
$\left<...\right>_V$ and $\left<...\right>_B$
indicate averaging over bias voltage and magnetic field, respectively.
The half width at half maximum gives us the correlation field
$B_c$ which is the quasi period of the fluctuations.
The inset in Fig. \ref{stat1} (a) shows that the correlation field
$B_c\approx 0.1$ T is rather independent of the magnetic-field orientation
$\theta$ with respect to the direction of the current. This independence
proves the 3D origin of the fluctuation pattern and, thus, demonstrates the
3D character of the quantum states in the emitter contact.
The Fourier transform of $\left<K_{\Delta B}\right>_V$
is the power spectrum $\left<S_B\right>_V$ of the
fluctuations plotted in Fig. \ref{stat1} (b).
It decays exponentially over more than two orders of magnitude.
A fit according to $\left<S_B\right>_V\propto\exp (-f_BB_c)$ yields
again $B_c\approx 0.1$ T.
The autocorrelation function as a function of bias voltage
$\left<K_{\Delta V}\right>_B$ and the corresponding power spectrum
$\left<S_V\right>_B$ are plotted in Figs. \ref{stat1} (c) and (d).
The correlation voltage is $V_c\approx 0.3$ mV
as taken from the autocorrelation function and
$V_c\approx 0.5$ mV as obtained from an exponential fit to the
power spectrum. This corresponds to a correlation energy of
$E_c=\alpha eV_c\approx 0.2$~meV.

Figure \ref{stat2} (a) shows the correlation field $B_c$ as a
function of bias voltage extracted from $K_{\Delta B}$
which was computed using individual magnetic-field traces of the
differential conductance.
The corresponding dependence of the correlation voltage
$V_c$ on magnetic field extracted from
$K_{\Delta V}$ is plotted in Fig. \ref{stat2} (b).
The correlation field varies irregularly on the scale of $V_c$ while the
correlation voltage changes on the scale of $B_c$, which confirms the random
character of the fine structure. The mean square of the fluctuations
exhibits a similar behavior: Figures \ref{stat2} (c) and (d) show that
var$_BG$ changes on the scale of $V_c$, while var$_VG$ varies on
the scale of $B_c$.

As the magnitude of the LDOS fluctuations is
given by the magnitude of the current fluctuations, we plot in
Figs. \ref{stat3} (a) and (b) the normalized variance of the current
as a function of bias voltage and magnetic field.
The quantities var$_BI/I^2$ and var$_VI/I^2$ vary on a larger scale
than var$_BG$ and var$_VG$ since they contain contributions of
long-range fluctuations which are suppressed in the differential conductance.
The autocorrelation functions and power spectra of the
current data yield $B_c\approx 0.2$ T and $E_c\approx 0.4$~meV.
{}From Fig. \ref{stat3}, the normalized variance of the LDOS fluctuations
is determined to var$\nu/\nu^2\propto$ var$I/I^2\approx~0.5\times 10^{-2}$.

We discuss these results using the semiclassical picture of quantum
interference between scattered electron waves.
At zero magnetic field, the local fluctuation pattern of the density of
states can be imagined as a superposition of the tails of Friedel
oscillations extended from the closest impurities \cite{pin66}.
The magnetic field dependence of the LDOS fluctuations is formed by
closed paths able to encircle magnetic flux as shown by the sketch in
Fig. \ref{stat3} \cite{field}.
The above determined correlation magnetic field agrees roughly
with the value of $B_c\sim\phi_0/l^2\approx 0.4$ T
calculated with $\phi_0$ the magnetic flux
quantum and $l\approx 100$ nm the mean free path deduced from the doping
density of the disordered contact.
Thus, we conclude that the observed fluctuations are formed by
semiclassical trajectories on the length scale of the mean free path,
similar to weak localization corrections to the conductivity in 3D systems.
It is however important to note that the LDOS exhibits no weak localization
correction, i.e. the tunneling current and the magnitude of the
main conductance resonances do not increase with
growing magnetic field, see Fig. \ref{cult}.
The correlation energy of LDOS fluctuations generated on the length scale
of the mean free path is theoretically expected to be
$E_c\sim\hbar/\tau=\hbar(v/l)$ with $\tau$ the momentum
relaxation time and $v$ the Fermi velocity. From the sample parameters we
estimate $E_c\sim 1$ meV which is of the same order of magnitude as
experimentally observed.
The theoretical expectation \cite{ler88} for the amplitude of the
LDOS fluctuations in 3D systems is var$\nu /\nu^2\sim (\lambda /l)^2$
with $\lambda$
being the Fermi wavelength. In our case a value of $(\lambda /l)^2\sim
10^{-2}$ results in reasonable agreement with experiment.
Finally, we want to point out that var$_VG$ exhibits
a maximum at $B=0$ T (Fig. \ref{stat2} (d))
which is expected for the magnitude of LDOS
fluctuations due to breaking of the time-reversal symmetry in
magnetic field.

In summary, we studied resonant tunneling through a discrete
localized level in a semiconductor double-barrier
heterostructure at low magnetic fields.
Mesoscopic fluctuations of the local density of states in the
emitter contact are observed as temperature insensitive
fine structure in the tunneling current. The analysis of
correlation functions and magnitude of the fine structure demonstrates,
that the fluctuation pattern is generated by quantum interference of
scattered electronic orbits at the length scale of the
mean free path of the disordered bulk semiconductor.

We thank M. Tewordt and B. Sch\"onherr for expert help during the
sample fabrication and A. K. Geim and I. V. Lerner for helpful discussions.
This work has been supported by the Bundesministerium
f\"ur Bildung, Wissenschaft, Forschung und Technologie.

\begin{figure}
\caption{(a) Current-voltage charcteristics $I(V)$ measured at
$T\approx 30$ mK (a). (b) Oscillations of the differential
conductance $G=dI/dV$ recorded at $T\approx 30$ mK (solid) as well as
$T=1.0$ K (dotted). The schematical drawing on top
illustrates the spectroscopy of the local density of states in the
disordered emitter contact of a double-barrier heterostructure.
\label{sketch}}
\end{figure}
\begin{figure}
\caption{Differential conductance vs bias voltage and magnetic field
(step 0.01 T) as surface plot (a) and color map (b) (white,
$G\leq -0.08$ $\mu$S; black, $G\geq 0.10$ $\mu$S).
\label{cult}}
\end{figure}
\begin{figure}
\caption{Voltage-averaged autocorrelation function (a) and
power spectral density (b) in magnetic field;
field-averaged autocorrelation function (c)
and power spectral density (d) in voltage ($V=25-30$ mV
and $B=0-2$ T). The inset of (a) shows the correlation field
as function of the magnetic-field orientation $\theta$ with respect to the
direction of the current.
The solid lines in (b) and (d) are exponential fits to the power spectra.
\label{stat1}}
\end{figure}
\begin{figure}
\caption{Correlation field vs bias voltage (a) and correlation voltage vs
magnetic field (b); variance of the differential conductance
var$_BG=\left<\delta G^2(B)\right>_B$ vs bias voltage (c) and
var$_VG=\left<\delta G^2(V)\right>_V$ vs magnetic field (d).
\label{stat2}}
\end{figure}
\begin{figure}
\caption{Normalized variance of the current var$_BI/I^2$
vs bias voltage (a) and var$_VI/I^2$ vs magnetic field (b).
Average values are given by dashed lines.
The sketch on top visualizes a typical closed electron trajectory
responsible for fluctuations of the local density of states.
\label{stat3}}
\end{figure}


\begin{references}

\bibitem[*]{h}New address: Institut f\"ur Festk\"orperphysik, Universit\"at
  Hannover, Appelstr. 2, 30167 Hannover, Germany.

\bibitem[\dag]{f}New address: Department of Theoretical Physics,
Oxford University, OX1 3NP Oxford, UK.

\bibitem{ber84}G. Bergmann, Phys. Rep. {\bf 107}, 1 (1984);
P. A. Lee and T. V. Ramakrishnan, Rev. Mod. Phys. {\bf 57}, 287 (1985).

\bibitem{was86}S. Washburn and R. A. Webb, Adv. Phys. {\bf 35}, 375 (1986).

\bibitem{alt91}For a review see
{\em Mesoscopic Phenomena in Solids}, edited by B. L. Altshuler,
P. A. Lee and R. A. Webb (North-Holland, Amsterdam, 1991).

\bibitem{alt86}B. L. Altshuler and B. I.
Shklovskii, Zh. Eksp. Teor. Fiz. {\bf 91}, 220
(1986) [Sov. Phys. JETP {\bf 64}, 127 (1986)].

\bibitem{ler88}I. V. Lerner, Phys. Lett. A {\bf 133}, 253 (1988);
B. L. Altshuler, V. E. Kravtsov and I. V. Lerner, p. 449 in Ref. \cite{alt91}.

\bibitem{cro93}M. F. Crommie, C. P. Lutz and D. M. Eigler,
Nature {\bf 363}, 524 (1993); Science {\bf 262}, 218 (1993);
E. J. Heller, M. F. Crommie, C. P. Lutz and D. M. Eigler, Nature {\bf 369},
464 (1994).

\bibitem{siv94}U. Sivan, F. P. Milliken, K. Milkove, S. Rishton, Y. Lee,
J. M. Hong, V. Boegli, D. Kern and M. DeFranza, Europhys. Lett. {\bf 25},
605 (1994).

\bibitem{ler92}I. V. Lerner and M. E. Raikh, Phys. Rev. B {\bf 45},
14036 (1992).

\bibitem{del92}M. W. Dellow, P. H. Beton, C. J. G. M. Langerak, T. J. Foster,
P. C. Main, L. Eaves, M. Henini, S. P. Beaumont and C. D. W. Wilkinson,
Phys. Rev. Lett. {\bf 68}, 1754 (1992);
M. Tewordt, L. Mart\'{\i}n-Moreno, V. J. Law, M. J. Kelly, R. Newbury,
M. Pepper, D. A. Ritchie, J. E. F. Frost and G. A. C. Jones,
Phys. Rev. B {\bf 46}, 3951 (1992).

\bibitem{su92}B. Su, V. J. Goldman and J. E. Cunningham, Science {\bf 255},
313 (1992); Phys. Rev. B {\bf 46}, 7644 (1992).

\bibitem{gei94}A. K. Geim, P. C. Main, N. La Scala, Jr., L. Eaves,
T. J. Foster, P. H. Beton, J. W. Sakai, F. W. Sheard, and M. Henini,
Phys. Rev. Lett. {\bf 72}, 2061 (1994).

\bibitem{des94}M. R. Deshpande, E. S. Hornbeck, P. Kozodoy, N. H. Dekker,
J. W. Sleight, M. A. Reed, C. L. Fernando and W. R. Frensley,
Semicond. Sci. Technol. {\bf 9}, 1919 (1994).

\bibitem{mcd95}P. J. McDonnell, A. K. Geim, P. C. Main, T. J. Foster,
P. H. Beton and L. Eaves, Physica B {\bf 211}, 433 (1995).

\bibitem{sch95a}T. Schmidt, M. Tewordt, R. J. Haug, K. v. Klitzing,
A. F\"orster and H. L\"uth, Solid State Electron., in press.

\bibitem{pin66}D. Pines and P. Nozi\`eres, The Theory of Quantum Liquids
(Plenum, New York, 1966).

\bibitem{field}In our limit of low magnetic fields,
the cyclotron frequency is smaller than the
momentum relaxation rate of the electrons in the doped contact,
$\omega_c\tau\leq 1$. Hence, Landau quantization has not to be considered
and the semiclassical picture of scattered electron waves is applicable.
In the opposite limit of high magnetic fields, $\omega_c\tau\gg 1$,
the irregular fluctuation pattern analyzed in this paper
becomes more regular, as has been observed recently by different groups
(see M. R. Deshpande {\em et al.},
p. 1899 in {\em 22nd International Conference on the Physics
of Semiconductors}, edited by D. J. Lockwood
(World Scientific, Singapore, 1995),
as well as Refs. \cite{gei94} and \cite{sch95a}).

\end{references}
\end{document}